\documentclass[preprint,12pt]{elsarticle}
\usepackage{amsmath,amssymb,multirow,array,graphicx}
\usepackage{caption}
\usepackage{adjustbox}
\usepackage{natbib}
\biboptions{sort&compress}
\usepackage[margin=1.0in]{geometry}
\journal{Communications in Nonlinear Science and Numerical Simulation}

\begin{document}

\begin{frontmatter}
\title {Systematic designing of bi-rhythmic and tri-rhythmic models in families of Van der Pol and Rayleigh oscillators}

\author{Sandip Saha, Gautam Gangopadhyay}
\address{S N Bose National Centre For Basic Sciences,\\ 
Block-JD, Sector-III, Salt Lake, Kolkata-700106, India}


\author{Deb Shankar Ray}
\address{Indian Association for the Cultivation of Science,\\ 
Jadavpur, Kolkata 700032, India}

\date{\today}

\begin{abstract}
Van der Pol and Rayleigh oscillators are two traditional paradigms of nonlinear dynamics. They can be subsumed into a general form of Li\'enard--Levinson--Smith(LLS) system. Based on a recipe for finding out maximum number of limit cycles possible for a class of LLS oscillator, we propose here a scheme for systematic designing  of generalised Rayleigh and Van der Pol families of oscillators with a desired number of multiple limit cycles. Numerical simulations are explicitly carried out for systematic search of the parameter space for bi-rhythmic and tri-rhythmic systems and their higher order variants. 
\end{abstract}

\begin{keyword}
Li\'enard--Levinson--Smith oscillator, bi-rhythmic and  tri-rhythmic  oscillators, Van der Pol oscillator, Rayleigh oscillator
\end{keyword}

\end{frontmatter}

\section{Introduction}

 Dissipative nonlinear dynamical systems\cite{murraybio,strogatz,epstein,goldbook,remickens,slross,len0,perko} are often described by minimal models governed by autonomous coupled differential equations\cite{slross,len0} which admit of periodic orbits in the form of limit cycles in two-dimensional phase space\cite{strogatz,len0,perko}. This kind of self-excited periodic motion is generically distinct from the forced or parametric oscillation\cite{dsrrayleigh} and arises as an instability of motion when the dynamical system at a steady state is subjected to a small perturbation. The self-excited oscillation is well known in heart beat\cite{Babloyantz_1988,Jenkins_2013}, nerve impulse propagation through neurons\cite{FitzHugh1955,fhn2PRE2015}, circadian oscillation\cite{cirgoldbeter1995} or sleep-wake cycle, glycolytic oscillation\cite{gly1,gly2,gly3,limiso} in controlling metabolic activity in a living cell and many other biological phenomena\cite{murraybio,goldbook,kaiser83,kaiser91,k-dsr,k-y2007c,k-y2010,biswas_pre_2016,biswas_chaos_2017}. Several musical instruments and human voice in acoustics, lasers in radiation-matter interaction, electrical circuits involving nonlinear vacuum tubes, oscillatory chemical reactions\cite{epstein,brusselator2009,len4} concern self-excited oscillations. Apart from understanding these various phenomena in terms of the limit cycle solutions of the nonlinear differential equations, self-excited oscillations are also interesting from the point of view of energetics, as they are not induced by any external periodic forcing\cite{dsrrayleigh,penvo}; rather the oscillation itself controls the driving force to act in phase with velocity. This results in a negative damping that acts as an energy source of oscillation. From the perspective of energetics or thermodynamics, as emphasized in Ref.~\cite{Jenkins_2013}, it is useful to regard the mathematical limit cycle as representing a ``thermodynamic cycle" in which the thermodynamic state of the system varies in time but repeats over a cycle after a finite period. This requires an active non-conservative force (effectively the anti-damping) as well as a load (the nonlinear damping) in the dynamical system. This aspect of self-oscillation has been successfully utilized to revisit the concept of thermodynamic heat engine in a solar cell\cite{Alicki_2017} and in electron shuttle\cite{W_chtler_2019}, a model nanoscale system.

 Two traditional celebrated paradigms of self-excited oscillators are Van der Pol oscillator\cite{strogatz,len0,rand2012} and Rayleigh oscillator\cite{strogatz,jwstrutt,dsr_Gen_Rayleigh_2013}. It has been shown \cite{len4,dsr_Gen_Rayleigh_2013,limiso,lccounting} that for a wide class of kinetic models describing biological and chemical oscillations in two-dimensional phase space variables can be cast in the form of either Rayleigh or Van der Pol oscillator or any of their  generalizations\cite{lls1,lls2,lls3,remickens,len4,dsr_Gen_Rayleigh_2013,limiso,powerlaw,lccounting}. Both the oscillators, however, can be subsumed into a common form, i.e., Li\'enard--Levinson--Smith(LLS) oscillator\cite{lls1,lls2,lls3,remickens,len4,dsr_Gen_Rayleigh_2013,limiso,powerlaw,lccounting}, so that they can be viewed as the two special cases\cite{lccounting} of LLS system. While the standard Rayleigh or Van der Pol oscillator allows single limit cycle, because of polynomial nature of nonlinear damping force and restoring force functions, LLS system exhibits multi-rhythmicity\cite{laurent1999multistability,goldbeter2019}, i.e., one observes the co-existence of multiple limit cycles in the dynamical system. In some biological systems nature utilizes this multi-rhythmicity as models  of regulation and in various auto-organisation of cell signalling\cite{goldbook,laurent1999multistability,	thomas2001multistationarity,goldbeter2018,goldbeter2019}. In a related issue a bi-rhythmic model for glycolytic oscillation was proposed by Decorly and Goldbeter\cite{decroly1982birhythmicity}. The coupling of two cellular oscillations\cite{goldbeter2019} also leads to multi-rhythmicity. By extending Van der Pol oscillator Kaiser had suggested a bi-rhythmic model\cite{kaiser83,kaiser91} which has subsequently been used in several occasions\cite{k-dsr,k-y2007c,k-y2010,biswas_pre_2016,biswas_chaos_2017,guo2019}.

 In spite of several interesting studies in different contexts as mentioned above, a systematic procedure for constructing a multi-rhythmic model with a desired number of limit cycles is still lacking. Two problems must be clearly distinguished at this juncture. The first one concerns of finding out the maximum number of limit cycles possible for a LLS system. The problem has been addressed in Ref.~\cite{countinglcjkb,infdampinglcjkb,lcbounestjkb} and also by us\cite{lccounting}. The second one, our focal theme in this paper is to systematically construct a minimal model with a desired number of limit cycles starting from a LLS system with a single limit cycle. The essential elements for this design is to choose the appropriate forms of polynomial damping and the restoring force functions. A scheme for critical estimation of the associated parameters for the polynomial functions and a smallness parameter is a necessary requirement. This non-trivial systematization of parameter space allows us to construct the higher order variants of both Van der Pol and Rayleigh oscillators with three, five and higher number of limit cycles. As illustration, we have proposed two cases of Van der Pol family; first one concerns five limit cycles of which three are stable and two are unstable dividing the basins of attractions. Second one is an alternative version of the bi-rhythmic Kaiser system\cite{kaiser83,kaiser91}. We have also proposed the bi- and tri-rhythmic models for the Rayleigh family of oscillators which are hitherto unknown to the best of our knowledge in the context of nonlinear oscillators. Our analysis shows while the mono-rhythmic Rayleigh and Van der Pol oscillators are unique, their bi- or tri-rhythmic or higher order variants may assume different forms depending on the nature of the polynomial functions. Our theoretical analysis is corroborated by detailed numerical simulations. 

\section{Polynomial damping and restoring force functions for Li\'enard--Levinson--Smith system; number of limit cycles}
\label{max-cycle}

We begin with a class of LLS equation of the following form
\begin{equation}
\ddot{\xi}+F(\xi,\dot{\xi}) \dot{\xi} +G(\xi)=0,
\label{eq:LienardForm}
\end{equation}
where, $F(\xi,\dot{\xi})$ and $G(\xi)$ being the polynomial functions as given by
\begin{align}
F(\xi,\dot{\xi})&=-[A_{01}+\sum_{n>0} A_{n1} \xi^n +\sum_{n\ge0} \sum_{m>1} A_{nm} \xi^n \dot{\xi}^{m-1}],   \nonumber\\
G(\xi) &=-[A_{10}+\sum_{n>1} A_{n0} \xi^{n-1} ] \xi.
\label{eq:LienardDampingRestoring}
\end{align}

They refer to the nonlinear damping and force functions, respectively. Depending  on the several conditions\cite{remickens,len4,lls1,lls2,lls3}on $F$ and $G$ for the existence of at least one locally stable limit cycle for dynamical model,  the following two cases can appear:

\textbf{I~:} When $A_{nm}=0$, $n\ge2, \forall m$, the above form of Eq.~\eqref{eq:LienardForm} takes the form, 
\begin{equation}
\ddot{\xi}+F_R(\dot{\xi}) \dot{\xi} +G_R(\dot{\xi})\xi=0,
\end{equation}
where, 
\begin{align}
F_R(\dot{\xi})&=-[A_{01}+\sum_{m>1} A_{0m} \dot{\xi}^{m-1}],\hspace{10 pt} G_R(\dot{\xi}) =-[A_{10}+\sum_{m>0} A_{1m} \dot{\xi}^{m}],
\end{align}
with the steady state ($\xi_s=0$) for a restoring force which is linear in $\xi$. It is a form of generalised Rayleigh oscillator\cite{dsr_Gen_Rayleigh_2013} where the condition of limit cycle reduces to, $F_R(0)<0$. One can have $F_R(\dot{\xi})=\epsilon (\dot{\xi}^2-1)$ and $G_R(\dot{\xi})=1$ for the special case of Rayleigh oscillator.

\textbf{II~:} When $A_{nm}=0$, $m\ge2,\forall n$, gives a Li\'enard equation with the steady state ($\xi_s=0$). The oscillator form can be written as 
\begin{equation}
\ddot{\xi}+F_L(\xi) \dot{\xi} +G_L(\xi)=0,
\end{equation}
with, 
\begin{align}
F_L(\xi)& =-[A_{01}+\sum_{n>0} A_{n1} \xi^n ],\hspace{10pt} G_L(\xi)=-[A_{10}+\sum_{n>1} A_{n0} \xi^{n-1} ] \xi,
\end{align}
where the condition of limit cycle is $F_L(0)<0$. It is to be noted that for a Li\'enard system  $F_L(\xi)$ and $G_L(\xi)$ are even and odd functions of $\xi$, respectively.  With the special case of Van der Pol oscillator we have $F_L(\xi)=\epsilon (\xi^2-1)$ and $G_L(\xi)=\xi.$

Eq.~\eqref{eq:LienardForm} is the starting point of our analysis. In a recent communication it has been shown that by implementing Krylov-Bogolyubov method \cite{slross,len0,lccounting,kbbook,rand2012} one can estimate the maximum number of limit cycles admissible by a dynamical system. Here the basic idea is to introduce the scaled time $\omega t \rightarrow \tau$ and the transformed variables $\xi \rightarrow Z, \dot{\xi} \rightarrow \omega \dot{Z}$ so that $Z(\tau)$ and $\dot{Z}(\tau)$ can be expressed as $Z(\tau) \approx r(\tau) ~ \cos (\tau+\phi(\tau))$ and $\dot{Z}(\tau) \approx -r(\tau) ~ \sin (\tau+\phi(\tau))$. Krylov-Bogolyubov averages leads us to the equations for average amplitude $\overline{r}$ and average phase $\overline{\phi}$. To proceed further we first terminate the series upto the value, $M,N$, as the highest power of $\dot{\xi}$ and $\xi$, respectively. For an  amplitude equation we take $M=N$ upto $3$. It covers all possible even and odd functions of $F(\xi,\dot{\xi})$ and $G(\xi)$, respectively. Such forms of $F(\xi,\dot{\xi})$ and $G(\xi)$ are  reduced to the forms given below,
\begin{align}
F(\xi,\dot{\xi}) &=-[A_{01}+A_{11} \xi+A_{21} \xi^2+A_{31} \xi^3 +A_{02} \dot{\xi}+A_{12} \xi \dot{\xi}+A_{22} \xi^2 \dot{\xi}+A_{32} \xi^3 \dot{\xi} \nonumber\\
&+A_{03} \dot{\xi}^{2}+A_{13} \xi \dot{\xi}^{2}+A_{23} \xi^2 \dot{\xi}^{2}+A_{33} \xi^3 \dot{\xi}^{2}], \nonumber\\
G(\xi) &=-[A_{10} \xi+A_{20} \xi^{2}+A_{30} \xi^{3}].
\end{align}
Abbreviating  $|F(0,0)|=\sigma \in \mathbb{R}^+$, with $F(\xi,\dot{\xi})=\sigma F_{\sigma} (\xi,\dot{\xi})$, the LLS equation after rescaling can be rewritten as
\begin{align}
\ddot{Z}(\tau)+\epsilon~h(Z(\tau),\dot{Z}(\tau))+Z(\tau)&=0,
\end{align}
where, $0<\epsilon=\frac{\sigma}{\omega^2}\ll 1$, 
$\omega^2=-A_{10}>0$ and $Z(\tau) \equiv \xi(t)$ with $\omega \dot{Z}(\tau) \equiv \dot{\xi}(t)$ and $h$ can be expressed as 
\begin{align}
h(Z,\dot{Z})=-\left[\lbrace H_1 +H_2 + H_3 \rbrace \omega \dot{Z}+ B_{20} Z^{2}+B_{30} Z^{3}\right], 
\end{align}
where, $H_1=B_{01}+ B_{11} Z+B_{21} Z^2+B_{31} Z^3,~ H_2=B_{02} \omega \dot{Z} + B_{12} Z \omega \dot{Z}+ B_{22} Z^2 \omega \dot{Z} + B_{32} Z^3 \omega \dot{Z},~ H_3=B_{03} \omega^2 \dot{Z}^{2}+B_{13} Z \omega^2 \dot{Z}^{2}+ B_{23} Z^2 \omega^2 \dot{Z}^{2}+B_{33} Z^3 \omega^2 \dot{Z}^{2}$ and $B_{ij}=\frac{A_{ij}}{\sigma}$, $i,j=0,1,2,3$ are the corresponding indices with $B_{0,0}=0$ and $B_{01}$  takes the  values (-1, 0, 1) depending on the property of the fixed point (asymptotically stable, center, limit cycle), respectively. Krylov-Bogolyubov averaging yields the following amplitude and phase equations,
\begin{align}
\dot{\overline{r}} &= \frac{\epsilon  \omega \overline{r} }{16} \lbrace\overline{r}^2 \left(B_{23} \overline{r}^2 \omega ^2+6 B_{03} \omega ^2+2 B_{21}\right)+8 B_{01}\rbrace+O(\epsilon^2), \nonumber\\
\dot{\overline{\phi}} &= -\frac{\epsilon \overline{r}^2}{16} \left(B_{32} \overline{r}^2 \omega ^2+2 B_{12} \omega ^2+6 B_{30}\right)+O(\epsilon^2).
\end{align}

A close scrutiny reveals that the effect on $\dot{\overline{r}}$ arises only in terms of the even coefficients of $F(\xi,\dot{\xi})$ in first order approximation. Again, only odd polynomial $G(\xi)$ plays a role in phase equation. One thus finds  at most four non-zero values of $\overline{r}$ having three distinct possibilities: (a) two sets of complex conjugate roots with asymptotically stable solution, (b) one pair of complex conjugate roots along with two real roots of equal magnitude having opposite sign giving a limit cycle solution and (c) either four real roots of equal magnitude of double multiplicity with opposite sign gives a limit cycle solution or two different sets of real roots of equal magnitude having opposite sign will provide limit cycle solutions of different radius. 

The roots of $\overline{r}$ with unique zero value gives a center \cite{powerlaw,len3.5}. The stability of the cycles thus can be examined by the $-ve$ or $+ve$ sign of $\frac{d \dot{\overline{r}}}{d \overline{r}}|_{\overline{r}=\overline{r}_{ss}}$ where $\frac{d \dot{\overline{r}}}{d\overline{r}}|_{\overline{r}=\overline{r}_{ss}=0}>0$ or $<0$ which determines the nature of the fixed point. Note that the limit cycle condition, $F(0,0)<0$ fails to give any clue about the number of limit cycles a system can admit. According to the root finding algorithm one can guess the maximum number of limit cycles of a LLS system\cite{lccounting,countinglcjkb,infdampinglcjkb}.

On a more general footing we have three possible combinations for having the maximum number of limit cycles or distinct non-zero real roots, which are given below:\\~\\
Case-I~: For both even or odd $N$ and $M$ there exist maximum $\frac{N+M}{2}-1$ limit cycles.\\~\\
Case-II~: For even $N$ and odd $M$ there exist maximum $\frac{N+M-1}{2}$ limit cycles.\\~\\
Case-III~: For odd $N$ and even $M$ there exist maximum $\frac{N+M-3}{2}$ limit cycles.\\

Determination of maximum number of possible limit cycles does not necessarily ensure the explicit functional form of the polynomials for damping and restoring forces, since the magnitude of the coefficients remain unknown. An important step in this direction is the systematic estimation of the parameter values or the coefficients of the polynomials for practical realisation of the oscillator with multiple limit cycles. In the next two sections we proceed to deal with this issue.

\section{On the generalisation of single-cycle oscillator to multi-cycle cases}

We now introduce the models for multi-cycle cases for Van der Pol and Rayleigh family of oscillators for k-cycles. In view of the above approach, we begin  examining some  physical models by classifying them according to single, two, three and k cycles for both families of oscillators which are available in the literature. The connection is discussed for the general model system with reference to the polynomial form of damping and restoring forces.

\subsection{Van der Pol family of cycles}

For Van der Pol family of oscillators with k-cycles\cite{perko,blows,lins} the equation in the LLS  form is
\begin{align}
\ddot{x}+(a_1+a_2 x+ \dots + a_{2k+1} x^{2k})\dot{x}+x&=0.
\end{align}

Considering the special case of Van der Pol oscillator developed by equation, $\ddot{x}+\epsilon (x^2-1) \dot{x}+x=0$ for weak nonlinearity with $ 0 < \epsilon \ll 1$, we obtain a stable limit cycle with $F(0,0)<0$. By referring to the earlier case-II as considered in the last section we have  $N=2$ and $M=1$, which gives a unique stable limit cycle. 

Two cycle cases are relatively rare. We rewrite the Li\'enard equation in the standard form as, $\dot{x(t)}=y(t)-F(x(t)), \dot{y(t)}=-x(t)$ with $F(x(t))$ being an odd polynomial. Subsequently we have, $\ddot{x}+F'(x)\dot{x}+x=0$ with $F'(x)=\frac{\partial F(x)}{\partial x}$ which now becomes an even polynomial. For $F(x)=a_1 x+a_2 x^2+a_3 x^3$, it has been shown\cite{perko,giacomini,blows,lins} that the system has a unique limit cycle if $a_1 a_3 <0$, which will be stable if $a_1 < 0$ and unstable if $a_1 > 0$. It corresponds to case-II for $N=2,M=1$. It is further extended by Rychkov\cite{rychkov} and showed that for $F(x)=(a_1 x + a_3 x^3 + a_5 x^5)$ the number of limit cycles is at most two. This observation corroborates with the numerical simulation when $F(x)$ is chosen\cite{perko,giacomini,blows,lins} as,  $F(x)=0.32 x^5 - \frac{4}{3} x^3 + 0.8 x$. Here the inner limit cycle is an unstable one as $F(0,0)=0.8>0$ but the outer one is stable. This correspond to case-II with $N=4$ and $M=1$. 

Extended Van der Pol oscillator for three cycle case is coined by Kaiser\cite{kaiser83,kaiser91,k-dsr,k-y2007c,k-y2010,k-y2007a} 
which is described as bi-rhythmicity, having the form, 
\begin{align}
\ddot{x}-\mu (1-x^2+\alpha x^4-\beta x^6) \dot{x}+x=0. \label{kaisereq}
\end{align}

Here the parameters $\alpha, \beta, \mu > 0$ of the  model provides an extremely rich variety of bifurcation phenomena and it exhibits bi-rhythmicity. The model allows three limit cycles with two stable cycle and in between the two stable limit cycles there will be an unstable one dividing the basins of attraction. In presence of an external electric field for the driven Kaiser model, the system shows some interesting phenomena\cite{kaiser83,kaiser91,k-dsr,k-y2007c,k-y2010,k-y2007a}. This model corresponds to the case-II with $N=6$ and $ M=1$. Therefore six roots may arise from the amplitude equation for $\mu>0$ and the controlling parameters, $(\alpha,\beta)$ have to be chosen from the zone of three limit cycles. Furthermore for $\beta=0$, the undriven Kaiser model with $\alpha=0.1$ gives two limit cycles  of which the smaller one will be stable and the larger one will be unstable. For LLS system, Blows and Lloyd\cite{perko,blows} have showed that, $\dot{x}=y-F(x), \dot{y}=-g(x)$ with $g(x)=x$ and $F(x)=a_1 x+a_2 x^2+ \dots + a_{2k+1} x^{2k+1}$ has at most $k$ limit cycles for the cases where the coefficients, $a_1,a_3,\dots,a_{2k+1}$ alternates in sign. This corresponds to the case for $N=2k$ and $M=1$ as the  condition of at most $k$ limit cycles. 

\subsection{Rayleigh family of cycles}

For Rayleigh family of oscillators with k-cycles\cite{gaiko2008} the equation may be written in  LLS  form as ,
\begin{align}
\ddot{x}- \left(\mu_1+\mu_2 \dot{x}+\mu_3 \dot{x}^2+ \dots +\mu_{2 k} \dot{x}^{2k-1}+\mu_{2 k+1} \dot{x}^{2 k} \right)\dot{x}+x &=0,
\label{gaiko}
\end{align}
Furthermore, considering the special case of ordinary Rayleigh oscillator\cite{jwstrutt,dsr_Gen_Rayleigh_2013},
\begin{align*}
\ddot{x}+\left( \eta_1 \dot{x}^2-\eta_2 \right)\dot{x}+\omega^2 x&=0,\quad \eta_1,\eta_2,\omega>0,
\end{align*}
one obtains an unique stable limit cycle for $\eta_2>0$. It corresponds to the case-I with  $N=1$ and $M=3$.

\emph{Multiple limit cycles}\cite{perko,gaiko2008,giacomini} \emph{in Rayleigh family are not known in literature}. Gaiko\cite{gaiko2008} has shown through geometrical approach that for a Li\'enard-type system having the form of Eq.~\eqref{gaiko}, one can have at most $k$ limit cycles iff $\mu_1>0$ and \emph{no physical example is available}. This result\cite{gaiko2008} is nicely matched\cite{lccounting} and for any value of $k \in \mathbb{Z}$ it corresponds to the odd-odd subcase (see case-I) with $M=2k+1$ and $N=1$ which subsequently gives the at most $\frac{N+M}{2}-1=k$ number of cycles.

\section{Construction of new families of Van der Pol and Rayleigh oscillators with multiple limit cycles}

Having discussed the bi-rhythmic Van der Pol oscillator, i.e, the Kaiser model, we now explore tri-rhythmic cases  as further generalizations. We begin with alternative  generalisation of Van der Pol system for bi-rhythmic and tri-rhythmic cases with higher powers of velocity variables. Similarly, bi-rhythmic and tri-rhythmic models are worked out for the family of Rayleigh oscillator. 

\subsection{Van der Pol family of oscillators}
\label{multi}

\subsubsection{Generalisation of Van der Pol system; three stable limit cycles}

We return to LLS oscillator and extend  the Kaiser model by keeping  a proper combination of $N$ and $M$ for case-II and an alternate sign condition on the coefficients of polynomial\cite{perko,blows}. An extended model beyond Kaiser that includes one more stable limit cycle can be  written as,
\begin{align}
\ddot{x}-\mu (1-x^2+\alpha x^4-\beta x^6+\gamma x^8-\delta x^{10}) \dot{x}+x=0,~\,\ 0< \mu \ll 1.
\label{K_Trirhm_xdot1_sys}
\end{align}

Here $ \mu $ is the Hopf bifurcation parameter as in Van der Pol model with the system parameters, $\alpha,\beta,\gamma,\delta>0$. According to case-II the maximum number of limit cycles the system can give is $\frac{10+1-1}{2}=5$. Further extension of Kaiser model leaving the parameter space $(\alpha,\beta)$ remain  unchanged, we find the parameter space for $(\gamma,\delta)$ which admits of three stable limit cycles. The associated amplitude equation for Eq.~\eqref{K_Trirhm_xdot1_sys}, is given by,
\begin{align}
\dot{\overline{r}} &=\frac{\overline{r} \mu}{1024}  \left(-21 \delta  \overline{r}^{10}+28 \gamma  \overline{r}^8-40 \beta  \overline{r}^6+64 \alpha  \overline{r}^4-128 \overline{r}^2+512\right).
\label{K_Trirhm _xdot1_req}
\end{align}

\begin{figure}[h!]
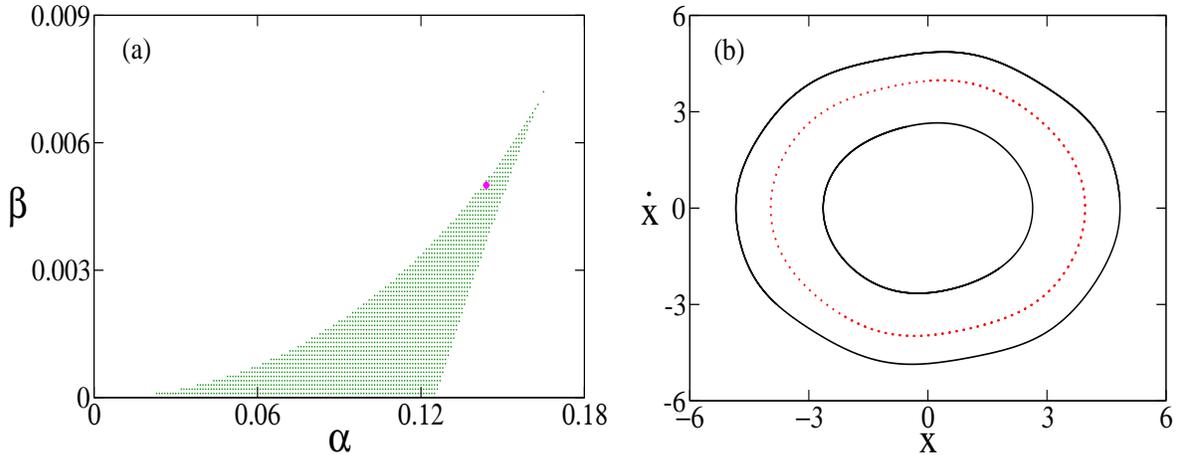

{\centering \vspace{1 cm}
\begin{tabular}{l c c }
\includegraphics*[height=6cm, width=8cm]{f1a-Reg_Pl_K_Birhm_xdot1.eps} &
\includegraphics*[height=6cm, width=7.1cm]{f1b-PS_Pl_K_Birhm_xdot1_eps_p1.eps}
\end{tabular}
\caption{\emph{Bi-rhythmic Van der Pol or Kaiser model} (Eq.~\ref{K_Trirhm_xdot1_sys}~; $\gamma=\delta=0$ with $\mu=0.1$). Subplot (a) represents the bi-rhythmic parameter space for $(\alpha,\beta)$ and (b) refer to the corresponding phase space plot showing the location of the stable limit cycles (black, continuous) along with an unstable limit cycle (red, dotted) for the parameters values, $\alpha=0.144~\text{ and}~\beta=0.005$ (magenta dot in subplot a).}
\label{K_Birhm_xdot1_fig} 
}
\end{figure}

For $\gamma=\delta=0$ amplitude equation reduces to that for Kaiser model. The corresponding parameter space for $(\alpha,\beta)$ and phase portrait of bi-rhythmicity (having amplitudes $\overline{r}_{ss}=2.63902,~3.96164~\text{and}~4.83953$) of Kaiser model are shown in Fig.~\ref{K_Birhm_xdot1_fig}.

\begin{figure}[h!]
{\centering \vspace{1 cm}
\begin{tabular}{l c c }
\includegraphics*[height=5.861cm, width=8cm]{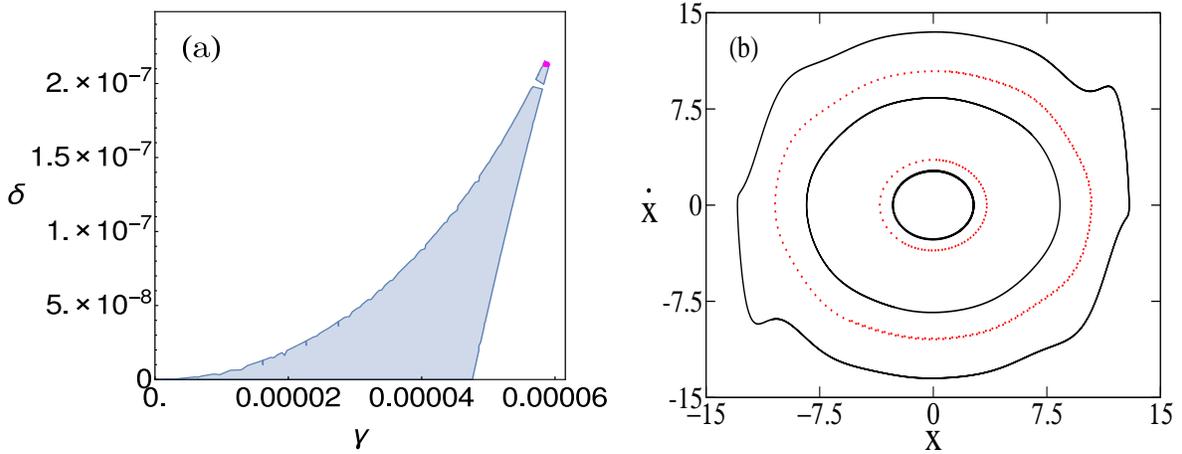} &
\includegraphics*[height=6cm, width=7.1cm]{f2b-PS_Pl_K_Trirhm_xdot1_eps_p01.eps}
\end{tabular}
\caption{\emph{Tri-rhythmic Van der Pol model} (Eq.~\ref{K_Trirhm_xdot1_sys}~; $\mu=0.01$). Subplot (a) represents the tri-rhythmic parameter space of $(\gamma,\delta)$ and (b) refer to the corresponding phase space plot of five concentric limit cycles among them three are stable (black, continuous) and remaining two are unstable (red, dotted) for the parameter values $\alpha=0.144,~\beta=0.005,~\gamma=0.00005862~\text{and}~\delta=2.13 \times 10^{-7}$ (magenta dots in subplot a and Fig.~\ref{K_Birhm_xdot1_fig}a).}
\label{K_Trirhm_xdot1_fig} 
}
\end{figure}

To realize tri-rhythmicity in Van der Pol system we search for the specific region  of $(\gamma,\delta)$ for a chosen set of values of  $(\alpha,\beta)$ corresponding to the bi-rhythmic space at $(\alpha=0.144,\beta=0.005)$, respectively. By solving Eq.~\eqref{K_Trirhm _xdot1_req} for the five distinct real roots, one obtains the region as shown in Fig.~\ref{K_Trirhm_xdot1_fig}(a). As no direct simple method  is available for an equation with degree higher than 3, we take resort to numerical simulation using higher order root finding algorithm in Mathematica as the amplitude equation is a polynomial in $\overline{r}$ of degree $11$ which can be reduced to degree $6$, where one root is zero and other non-zero roots  appear in conjugate pairs. 

Now, for $\gamma$ and $\delta$ from the region in Fig.~\ref{K_Trirhm_xdot1_fig}(a), we  obtain five distinct magnitudes of non zero real roots of Eq.~\eqref{K_Trirhm _xdot1_req}. The number of limit cycles can be obtained from the magnitude of the radii. The unique zero value as a root provides the location of the fixed point as well as the stability of the fixed point of the system. For $\gamma =0.00005862$ and $\delta=2.13 \times 10^{-7}$, it gives $\overline{r}_{ss}=0,~ 2.66673,~ 3.53498,~ 8.3682,~ 10.4793,~ 12.9421$. The stability of the corresponding cycles can be determined by the signs of the real parts of the eigenvalues  appearing in the respective order as $( -,~ +,~ -,~ +,~ - )$ for $\overline{r}_{ss}= 2.66673,~ 3.53498,~ 8.3682,~ 10.4793,~ 12.9421$. This implies that the cycles are in the order of stable, unstable, stable, unstable, stable, respectively. The unique fixed point $\overline{r}_{ss}=0$ (i.e., the origin) is unstable as $\frac{d \dot{\overline{r}}}{d \overline{r}}|_{\overline{r}_{ss}=0}>0$. The phase space plot of Fig.~\ref{K_Trirhm_xdot1_fig}(b) corresponds to the  stable cycles (black, continuous) along with unstable cycles (red, doted). One has to set $\mu$ is very small\cite{perko,blows} as much as possible to get nearly a circular orbit. 

\subsubsection{Another generalization of Van der Pol system: three stable limit cycles}

We now examine the LLS oscillator model to suggest an alternative  extension of Van der Pol model, where the damping part contains $\dot{x}^3$ instead of $\dot{x}$ as follows,
\begin{align}
\ddot{x}-\mu (1-x^2+\alpha x^4-\beta x^6+\gamma x^8-\delta x^{10}) \dot{x}^3+x=0,~ 0< \mu \ll 1; ~ \alpha,\beta,\gamma,\delta>0.
\label{K_Trirhm_xdot3_sys}
\end{align}

This corresponds to case-II,  $N=10$ and $M=3$; therefore the system can have atmost $\frac{10+3-1}{2}=6$ limit cycles. The numerical examination however, reveals  that the actual number is $5$. The  justification is provided below. The corresponding amplitude equation takes the form, 
\begin{align}
\dot{\overline{r}} &=\frac{\overline{r}^3 \mu}{2048}  \left(-9 \delta  \overline{r}^{10}+14 \gamma  \overline{r}^8-24 \beta  \overline{r}^6+48 \alpha  \overline{r}^4-128 \overline{r}^2+768\right).
\label{K_Trirhm_xdot3_req}
\end{align}

\begin{figure}[h!]
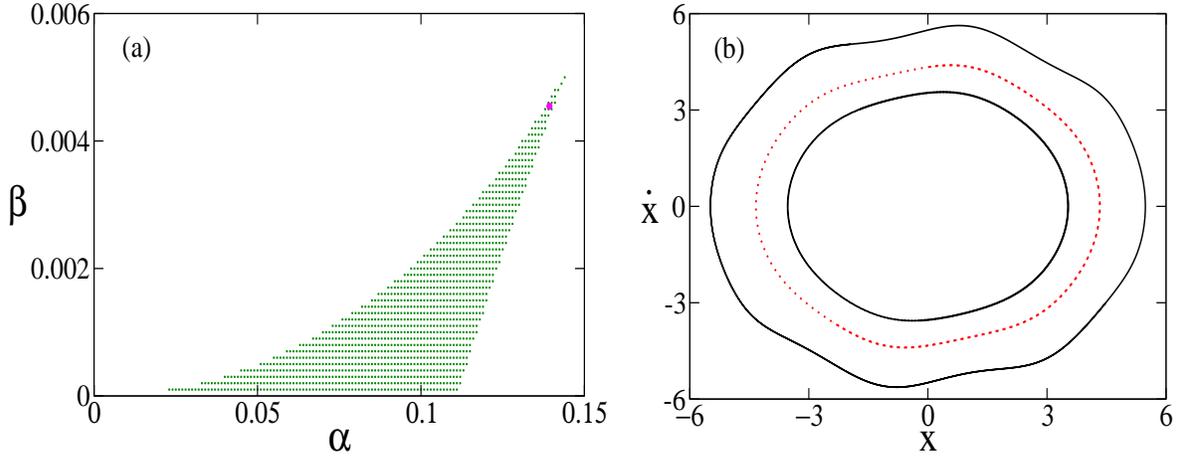

{\centering \vspace{1 cm}
\begin{tabular}{l c c }
\includegraphics*[height=6cm, width=8cm]{f3a-Reg_Pl_K_Birhm_xdot3.eps} &
\includegraphics*[height=6cm, width=7.1cm]{f3b-PS_Pl_K_Birhm_xdot3_eps_p01.eps}
\end{tabular}
\caption{\emph{A generalisation of bi-rhythmic Van der Pol oscillator} (Eq.~\ref{K_Trirhm_xdot3_sys}~; $\gamma=0=\delta$ with $\mu=0.01$). Subplot (a) represents the bi-rhythmic parameter space for $(\alpha,\beta)$ and (b) refer to the corresponding phase space plot showing the location of the stable limit cycles (black, continuous) along with the unstable limit cycle (red, dotted) for the parameters values, $\alpha=0.139317~\text{ and}~\beta=0.00454603$ (magenta dot in subplot a).}
\label{K_Birhm_xdot3_fig} 
}
\end{figure}

Now, to have a complete knowledge of the full parameter space for tri-rhythmicity for $(\alpha,\beta,\gamma,\delta)$--at first we have to fix $\gamma=\delta=0$, as in the  previous case. This is a generically distinct kind of variant of Kaiser model having three limit cycles. This type of bi-rhythmic oscillator is not known in the literature. For $\gamma=\delta=0$, case-II gives the maximum number of limit cycles $4$, but amplitude equation~\eqref{K_Trirhm_xdot3_req} says that it has at most three non-zero roots of distinct magnitudes and the zero root has multiplicity three. The bi-rhythmic parameter zone and the phase space plot for $\gamma=\delta=0$ are given in Fig.~\ref{K_Birhm_xdot3_fig}.

Now, for $\alpha=0.139317$ and $\beta=0.00454603$, one must have three non-zero values of $\overline{r}_{ss}= 3.53297,~4.33345,~5.48006$. The zero value of $\overline{r}_{ss}$ has multiplicity 3 which is basically a neutral fixed point but is unstable in nature. The stability of the cycles is given in an outward  sequence as stable, unstable and stable, respectively. 

\begin{figure}[h!]
{\centering \vspace{1 cm}
\begin{tabular}{l c c }
\includegraphics*[height=5.861cm, width=8cm]{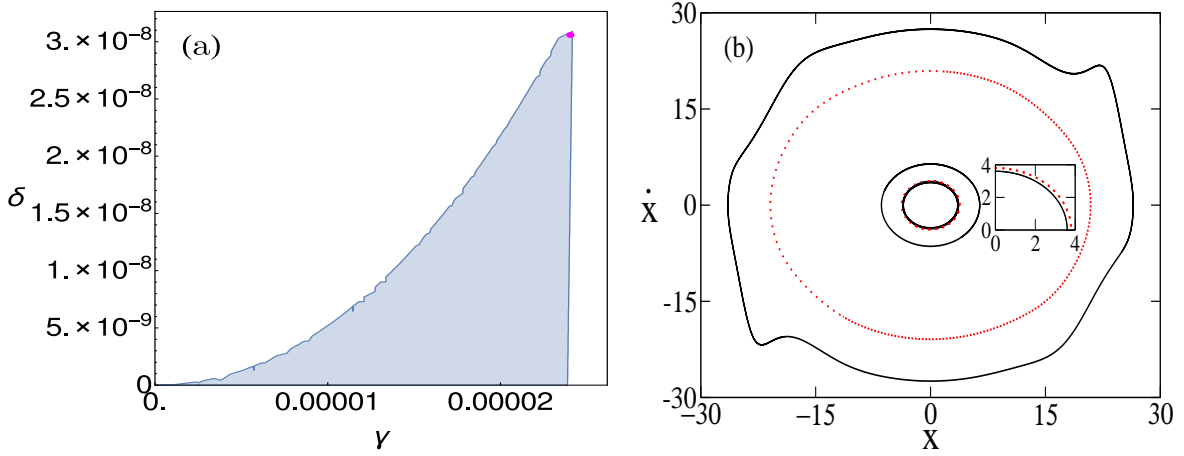} &
\includegraphics*[height=6cm, width=7.1cm]{f4b-PS_Pl_K_Trirhm_xdot3_eps_p0000001.eps}
\end{tabular}
\caption{\emph{Alternate generalisation of tri-rhythmic Van der Pol oscillator} (Eq.~\ref{K_Trirhm_xdot3_sys}~; $\mu=0.0000001$). Subplot (a) represents the tri-rhythmic parameter space of $(\gamma,\delta)$ and (b) refer to the corresponding phase space plot of five concentric limit cycles among them three are stable (black, continuous) and two are unstable (red, dotted) for the parameter values $\alpha=0.139317,~\beta=0.00454603,~\gamma=0.00002402~\text{and}~\delta=3.058 \times 10^{-8}$  (magenta dots in subplot a and Fig.~\ref{K_Birhm_xdot3_fig}a). The inset in subplot (b) zooms the gap between the first stable and unstable limit cycles in the outward direction.}
\label{K_Trirhm_xdot3_fig} 
}
\end{figure}

Now, to find the region of $(\gamma,\delta)$ in 2-D space, we first fix the parameters at $\alpha=0.139317~\text{and}~\beta=0.00454603$. Eq. \eqref{K_Trirhm_xdot3_req} is then solved  to provide five distinct real roots. The parameter region for $(\gamma,\delta)$ is given in Fig.~\ref{K_Trirhm_xdot3_fig}(a). For fixed values of $\gamma=0.00002402$ and $\delta=3.058 \times 10^{-8}$--the distinct roots of $\overline{r}_{ss}$ are $0$(of multiplicity 3), $3.75166,~ 3.80059,~ 6.63736,~ 20.9299,~ 26.6688$ and the non zero values are the radii of the cycles having the stability in outward sequence as: stable, unstable, stable, unstable, stable, respectively. The respective phase space portrait is given in Fig.~\ref{K_Trirhm_xdot3_fig}(b). 

\subsection{Rayleigh family of oscillators}

\subsubsection{Bi-rhythmic  Rayleigh: three stable limit cycles}

Here we consider an extension for Rayleigh oscillator model as a LLS system. The model can be written as,
\begin{align}
\ddot{x}-\mu (1-\dot{x}^2+\alpha \dot{x}^4-\beta \dot{x}^6+\gamma \dot{x}^8-\delta \dot{x}^{10}) \dot{x}+x=0,~ 0< \mu \ll 1;~ \alpha,\beta,\gamma,\delta>0.
\label{R_Trirhm_xdot1_sys}
\end{align}

Corresponding to case-I, we have $N=1$ and $M=11$ and accordingly  the system can have at most $\frac{11+1-2}{2}=5$ limit cycles. The amplitude equation takes the form,
\begin{align}
\dot{\overline{r}} &=\frac{\overline{r} \mu}{1024}  \left(-231 \delta  \overline{r}^{10}+252 \gamma  \overline{r}^8-280 \beta  \overline{r}^6+320 \alpha  \overline{r}^4-384 \overline{r}^2+512\right).
\label{R_Trirhm_xdot1_req}
\end{align}

\begin{figure}[h!]
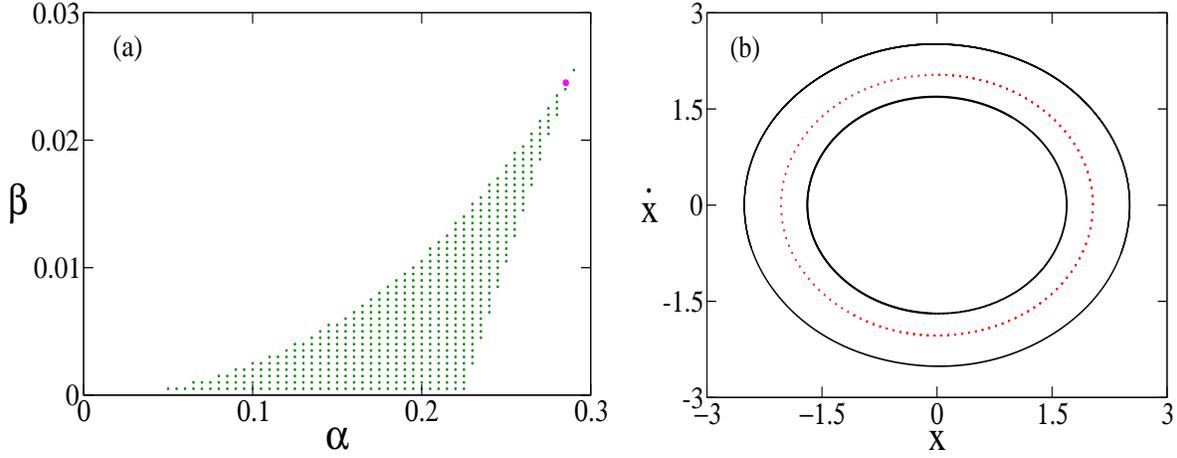

{\centering \vspace{1 cm}
\begin{tabular}{l c c }
\includegraphics*[height=6cm, width=8cm]{f5a-Reg_Pl_R_Birhm_xdot1.eps} &
\includegraphics*[height=6cm, width=7.1cm]{f5b-PS_Pl_R_Birhm_xdot1_eps_p1.eps}
\end{tabular}
\caption{\emph{Bi-rhythmic Rayleigh oscillator} (Eq.~\ref{R_Trirhm_xdot1_sys}~; $\gamma=0=\delta$ with $\mu=0.1$). Subplot (a) represents the bi-rhythmic parameter space for $(\alpha,\beta)$ and (b) refer to the corresponding phase space plot showing the location of the stable limit cycles (black, continuous) along with an unstable limit cycle (red, dotted) for the parameters values, $\alpha=0.285272~\text{ and}~\beta=0.0244993$ (magenta dot in subplot a).}
\label{R_Birhm_xdot1_fig}
}
\end{figure}

Proceeding as in the previous case we see that for $\gamma=\delta=0$ the  maximum  number of limit cycles is $3$ having the bi-rhythmic parameter region $(\alpha, \beta)$ as given in Fig.~\ref{R_Birhm_xdot1_fig}(a). The corresponding phase portrait is shown in Fig.~\ref{R_Birhm_xdot1_fig}(b) with amplitudes $\overline{r}_{ss}=~1.69091,~2.03334~\text{and}~ 2.51274$.

\begin{figure}[h!]
{\centering \vspace{1 cm}
\begin{tabular}{l c c }
\includegraphics*[height=5.861cm, width=8cm]{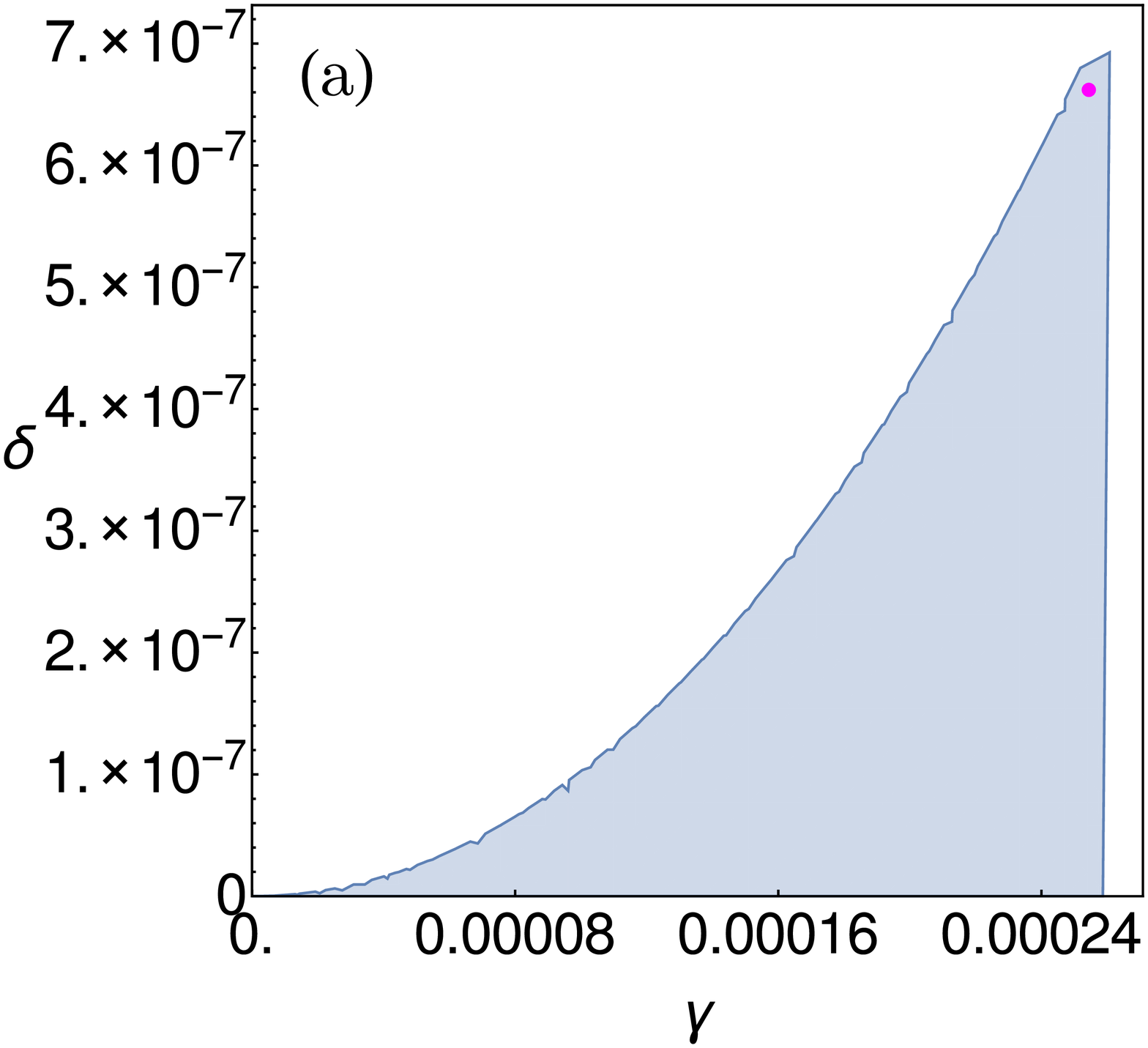} &
\includegraphics*[height=6cm, width=7.1cm]{f6b-PS_Pl_R_Trirhm_xdot1_eps_p00001.eps}
\end{tabular}
\caption{\emph{Tri-rhythmic Rayleigh oscillator} (Eq.~\ref{R_Trirhm_xdot1_sys}~; $\mu=0.00001$). Subplot (a) represents the tri-rhythmic parameter space of $(\gamma,\delta)$ and (b) refer the corresponding phase space plot of five concentric limit cycles among them three are stable (black, continuous) and two are unstable (red, dotted) for the parameter values $\alpha=0.285272,~\beta=0.0244993,~\gamma=0.0002544~\text{and}~\delta=6.62 \times 10^{-7}$ (magenta dots in subplot a and Fig.~\ref{R_Birhm_xdot1_fig}a). The inset in subplot (b) zooms the gap between the first stable and unstable limit cycles in the outward direction.}
\label{R_Trirhm_xdot1_fig} 
}
\end{figure}

For searching the region of $(\gamma,\delta)$ in 2-D space, $(\alpha,\beta)$ is fixed at $(0.285272,0.0244993).$ The parameter space $(\gamma,\delta)$ is shown in Fig.~\ref{R_Trirhm_xdot1_fig}(a). The phase portrait in Fig.~\ref{R_Trirhm_xdot1_fig}(b) shows the tri-rhythmicity for $\gamma=0.0002544~\text{and}~\delta=6.62 \times 10^{-7}$. The amplitudes for the tri-rhythmic Rayleigh oscillator, are , $\overline{r}_{ss}=~1.77779,~1.82091,~2.86779,~12.5239~ \text{and}~ 15.7377$--are in the same outward sequence i.e. stable, unstable, stable, unstable and stable.

\subsubsection{Alternative form of extended Rayleigh model: three stable limit cycles}

Here we consider another LLS oscillator system as an alternative form of the extended Rayleigh oscillator model with $N=1$ and $M=13$. The model can be written as,
\begin{align}
\ddot{x}-\mu (1-\dot{x}^2+\alpha \dot{x}^4-\beta \dot{x}^6+\gamma \dot{x}^8-\delta \dot{x}^{10}) \dot{x}^3+x=0,~ 0< \mu \ll 1;~ \alpha,\beta,\gamma,\delta>0.
\label{R_Trirhm_xdot3_sys}
\end{align}

As per case-I, the system has at most $\frac{13+1-2}{2}=6$ cycles and the corresponding amplitude equation takes the form, 
\begin{align}
\dot{\overline{r}} &=\frac{\mu  \overline{r}^3}{2048} \left(-429 \delta  \overline{r}^{10}+462 \gamma  \overline{r}^8-504 \beta  \overline{r}^6+560 \alpha  \overline{r}^4-640 \overline{r}^2+768\right).
\label{R_Trirhm_xdot3_req}
\end{align}

\begin{figure}[h!]
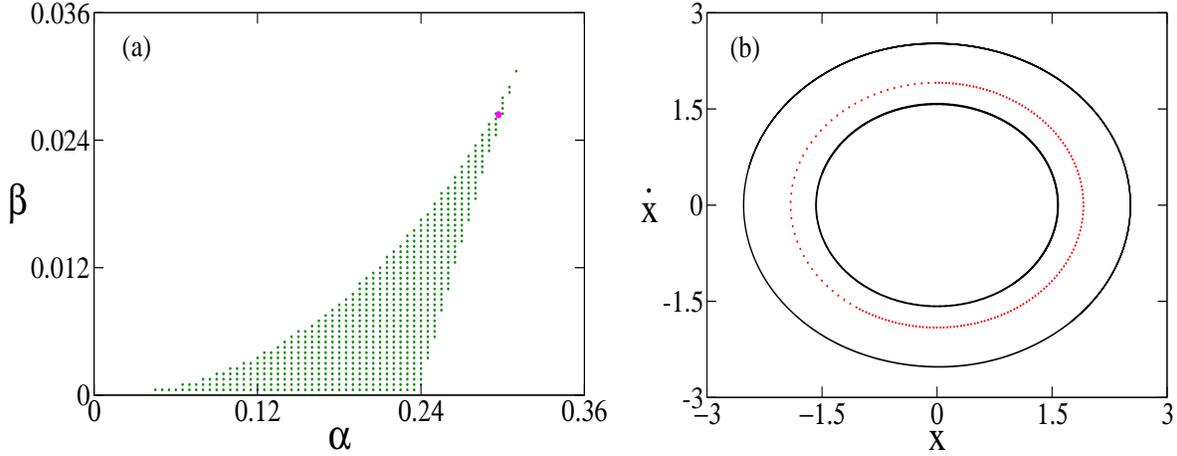

{\centering \vspace{1 cm}
\begin{tabular}{l c c }
\includegraphics*[height=6cm, width=8cm]{f7a-Reg_Pl_R_Birhm_xdot3.eps} &
\includegraphics*[height=6cm, width=7.1cm]{f7b-PS_Pl_R_Birhm_xdot3_eps_p01.eps}
\end{tabular}
\caption{\emph{Alternate generalisation of Rayleigh oscillator for Bi-rhythmicity} (Eq.~\ref{R_Trirhm_xdot3_sys}~; $\gamma=0=\delta$ with $\mu=0.01$). Subplot (a) represents the bi-rhythmic parameter space for $(\alpha,\beta)$ and (b) refer to the corresponding phase space plot showing the location of the stable limit cycles (black, continuous) along with an unstable limit cycle (red, dotted) for the parameters values, $\alpha=0.296930 ~\text{and}~ \beta=0.0264040$ (magenta dot in subplot a).}
\label{R_Birhm_xdot3_fig} 
}
\end{figure}

For  $\gamma=\delta=0$ we have at most $3$ limit cycles having the bi-rhythmic parameter region for $(\alpha,\beta)$ as shown in Fig.~\ref{R_Birhm_xdot3_fig}(a). The corresponding phase portrait is given in Fig.~\ref{R_Birhm_xdot3_fig}(b). Here, for the above bi-rhythmic Rayleigh oscillator the amplitudes will take the values, $\overline{r}_{ss}=1.5775,~1.90947~\text{and}~2.52202$.

\begin{figure}[h!]
{\centering \vspace{1 cm}
\begin{tabular}{l c c } 
\includegraphics*[height=5.861cm, width=8cm]{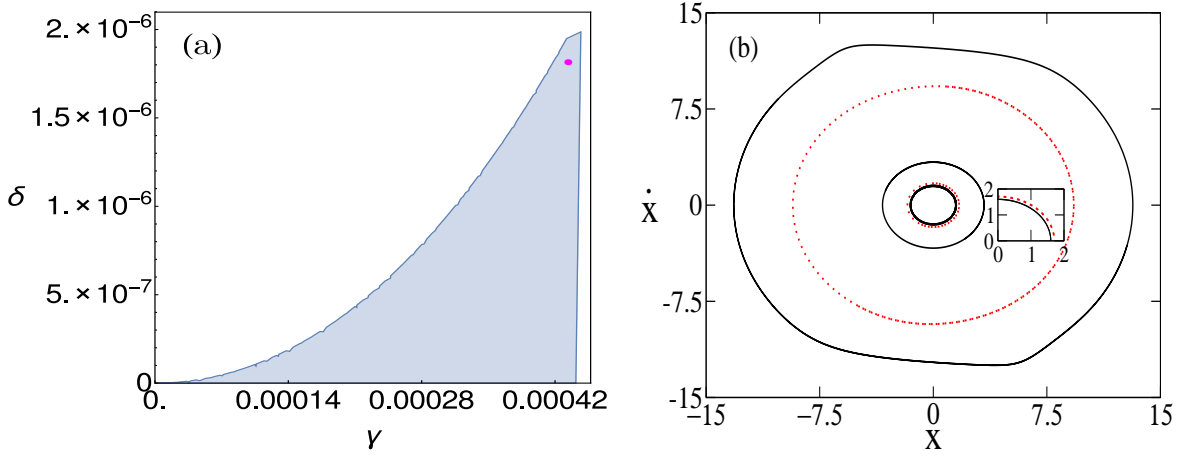} &
\includegraphics*[height=6cm, width=7.1cm]{f8b-PS_Pl_R_Trirhm_xdot3_eps_p0000005.eps}
\end{tabular}
\caption{\emph{Alternate generalisation of Rayleigh oscillator for tri-rhythmicity} (Eq.~\ref{R_Trirhm_xdot3_sys}~; $\mu=0.0000005$). Subplot (a) represents the tri-rhythmic parameter space of $(\gamma,\delta)$ and (b) refer to the corresponding phase space plot of five concentric limit cycles among them three are stable (black, continuous) and two are unstable (red, dotted) for the parameter values $\alpha=0.296930,~\beta=0.0264040,~\gamma=0.0004334~\text{and}~\delta=1.815 \times 10^{-6}$ (magenta dots in subplot a and Fig.~\ref{R_Birhm_xdot3_fig}a). The inset in subplot (b) zooms the gap between the first stable and unstable limit cycles in the outward direction.}
\label{R_Trirhm_xdot3_fig} 
}
\end{figure}

The parameter space of  $(\gamma,\delta)$ in 2-D space for a fixed  $\alpha=0.296930$ and $\beta=0.0264040$, is shown in Fig.~\ref{R_Trirhm_xdot3_fig}(a) and Fig.~\ref{R_Trirhm_xdot3_fig}(b) shows the corresponding tri-rhythmic phase portrait at a fixed values $(\gamma,~\delta)$ i.e., $\gamma=0.0004334~\text{and}~\delta=1.815 \times 10^{-6}$. The amplitudes of Eq.~\eqref{R_Trirhm_xdot3_req} will be the non zero steady states values of  $\overline{r}_{ss}$ i.e.,  $1.66034,~1.70743,~3.0214,~9.27171~\text{and}~12.5056$, which are in the same stability sequence as in previous.

By making use of this approach we have demonstrated the parameter spaces for bi-rhythmicity and tri-rhythmicity for Rayleigh and Van der Pol cases. The structure of the stable and unstable concentric multiple limit cycles are  described in terms of corresponding radial equations.   The scheme also covers the alternate generalizations of   bi-rhythmic and tri-rhythmic  Rayleigh and Van der Pol families of oscillators.

For the alternate cases (i.e., Eq.~\ref{K_Trirhm_xdot3_sys} and Eq.~\ref{R_Trirhm_xdot3_sys}), both the families of oscillators it appears that $F(0,0)=0$ which is a condition for center\cite{powerlaw}. However, both of the families have at least one stable limit cycle which can be checked from the amplitude equation, giving a nonzero radius. 
It can be resolved if we consider a small neighbourhood of $(0,0)$, say,  $(\delta_{1},\delta_{2})$ where $F(\delta_1,\delta_2)_{(\delta_{1},\delta_{2}) \rightarrow (0,0)} <0$.

\section{Summary, discussions and conclusions}

Based on a general scheme of counting limit cycles of a given LLS equation we have  proposed a recipe for systematically designing models of multi-rhythmicity. We note the basic tenets of the scheme stepwise as follows;\\~\\
Step-I: Given that the number of desired limit cycles is $k$, we may partition $k$ into $N$ and $M$ such that $k=(\frac{N+M}{2}-1)$ for Case-I; $k=(\frac{N+M-1}{2})$ for Case-II and $k=(\frac{N+M-3}{2})$ for Case-III taking care of even and odd nature of $N$ and $M$ as appropriate for Van der Pol or Rayleigh families of oscillators, $M$ and $N$ being the highest power of velocity $\dot{\xi}$ and position $\xi$, respectively for the damping function $F(\xi,\dot{\xi})$ and force function $G(\xi)$.\\~\\
Step-II: Once $N$ and $M$ are fixed we need to construct the polynomials for damping functions taking care of alternative signs and polynomial for the force function.\\~\\
Step-III: To fix the coefficients of the polynomials we first consider the lowest order polynomial functions that determine the mono or bi-rhythmic model. By choosing a point in the two dimensional parameter space for the specific nonzero values of the coefficients for the models with high multi-rhythmicity are switched on. For example, as in sec.~\ref{multi}, we choose one point from $\alpha-\beta$ parameter space for bi-rhythmic model and then proceed to construct $\gamma-\delta$ space (for a fixed $\alpha,~\beta$) for the model of tri-rhythmicity. The procedure may be repeated for higher order variants.\\~\\
Step-IV: A smallness parameter ($\mu$ as in Eq.~\ref{K_Trirhm_xdot1_sys}) can be introduced in the damping term which can be suitably tuned for numerical realisation of the high order limit cycles.\\~\\
In the table we have summarised several cases of oscillators belonging to Van der Pol and Rayleigh families of oscillators. In terms of the polynomial form of the damping and forcing functions of phase space variables the average amplitude and phase equations can be derived to obtain the maximum possible number of limit cycles of a dynamical system. It is verified that  for  a LLS system depending on the values of $N$ and $M$ which are the maximum degree of the polynomials for damping and restoring forces, respectively, the number of limit cycles can be at most $\frac{N+M-2}{2}$ (even-even or odd-odd sub cases) or  $\frac{N+M-1}{2}$ (even-odd  sub case) or $\frac{N+M-3}{2}$ (odd-even  sub case). The generalized  Li\'enard system can be recovered for $M=1$ and $N \in \mathbb{Z^+}$ whereas for the generalised Rayleigh oscillator we have, $N=1$ and $M \in \mathbb{Z^+}$. For polynomial form of damping and restoring force functions we have constructed bi-rhythmic and tri-rhythmic oscillators of Van der Pol and Rayleigh families. New alternative generalizations  of Van der Pol and Rayleigh families of oscillators are also introduced as models for bi-rhythmicity and tri-rhythmicity. Our approach shows that it is possible to construct a LLS system with arbitrary rhythmicity.

Secondly the scheme is used to determine the appropriate  range of parameters for realizing limit cycle oscillations as  shown by the corresponding phase portraits.  
Our approach shows that once the parameters space that allows a single limit cycle is determined, one may choose a suitable point in this space to select  only the parameters pertaining to the higher order terms of the polynomials for a systematic search for parameter space for constructing the remaining limit cycles. Since the choice of the point as referred to is not unique, it is imperative that the choice of parameters remains widely open for various generalisation of a multi-rhythmic system.

As multi-rhythmicity plays an important role in switching transitions between different dynamical states in a nonlinear system, its control\cite{Pisarchik2000} and manipulation\cite{k-y2007c, biswas_pre_2019} would be useful in many self-induced oscillatory processes\cite{strogatz,epstein,goldbook} in diverse interdisciplinary areas. The proposed multi-rhythmic oscillators can also be useful in various circuit designs as well as networks according to the demand of the system upon using different controlling schemes (e.g., delay-feedback control\cite{k-dsr}, self-conjugate feedback control\cite{biswas_pre_2016,biswas_chaos_2017} and a number of approaches reviewed by Pisarchik et al.\cite{Pisarchik2014}) to convert them into lower rhythmic systems.  
Systematic lowering of rhythmicity may also be utilized by suitably reducing the reaction rates of higher order polymeric reactions, using such schemes of control.

\newpage 
Nonlinear damping functions, classification of limit cycles of bi-rhythmic and tri-rhythmic cases of Van der Pol and Rayleigh families of oscillators

\hspace*{8 cm}
\begin{adjustbox}{angle=-270} \tiny{ 
\begin{tabular}{ |c|c|c|c|c|c| } 
&&\textbf{\large{Table }}&&&\\
\hline
\hline
&&&&&\\
$h(x,~\dot{x})$ & $f(r)$ & Stability & Regions \&  & Stability &    \\

&  & of FP  & fixed $(\alpha,~\beta)$  &  \& radii &  \\

&&&&&\\
\hline
\hline
&&&&&\\
$ - (1-x^2+\alpha x^4- \beta x^6) \dot{x}$  & $\overline{r}   \left(-5 \beta  \overline{r}^6+8 \alpha  \overline{r}^4-16 \overline{r}^2+64\right) $  & U & Fig.~\ref{K_Birhm_xdot1_fig}(a) & S,~U,~S & Bi\\
& & & $( 0.144,~0.005 )$ & 2.63902,~3.96164,~4.83953 & -r \\

&&&&&\\
$- (1-x^2+\alpha x^4-\beta x^6) \dot{x}^3$ & $\overline{r}^3  \left(-3 \beta  \overline{r}^6+6 \alpha  \overline{r}^4-16 \overline{r}^2+96\right) $  & NS & Fig.~\ref{K_Birhm_xdot3_fig}(a) & S,~U,~S & hy\\
&&& $( 0.139317,~0.00454603 )$ & 3.53297,~4.33345,~5.48006 & th\\

&&&&&\\
$- (1-\dot{x}^2+\alpha \dot{x}^4-\beta \dot{x}^6) \dot{x}$ & $\overline{r} \left(-35 \beta  \overline{r}^6+40 \alpha  \overline{r}^4-48 \overline{r}^2+64\right)  $  & U &  Fig.~\ref{R_Birhm_xdot1_fig}(a) & S,~U,~S  & mi\\
&&& $( 0.285272,~0.0244993 )$ & 1.69091,~2.03334,~2.51274  & c\\

&&&&&\\

$- (1-\dot{x}^2+\alpha \dot{x}^4-\beta \dot{x}^6) \dot{x}^3$ & $\overline{r}^3 \left(70 \alpha  \overline{r}^4-63 \beta  \overline{r}^6-80 \overline{r}^2+96\right) $  & NS &  Fig.~\ref{R_Birhm_xdot3_fig}(a) & S,~U,~S  & \\
&&& $(0.296930,~0.0264040 )$ & 1.5775,~1.90947,~2.52202 &\\
&&&&&\\
\hline

&&&&&\\
$ - (1-x^2+\alpha x^4- \beta x^6+\gamma x^8-  \delta x^{10}) \dot{x}$  & $\overline{r}   \left(-21 \delta  \overline{r}^{10}+28 \gamma  \overline{r}^8-40 \beta  \overline{r}^6+64 \alpha  \overline{r}^4-128 \overline{r}^2+512\right) $ & U & Fig.~\ref{K_Trirhm_xdot1_fig}(a)  & S,~U,~S,~U,~S & Tr\\
&&&  $(5.862 \times 10^{-5},~2.13 \times 10^{-7} )$ & 2.66673,~3.53498,~8.3682,~10.4793,~12.9421& i-\\

&&&&&\\
$- (1-x^2+\alpha x^4-\beta x^6+\gamma x^8-\delta x^{10}) \dot{x}^3$ & $\overline{r}^3  \left(-9 \delta  \overline{r}^{10}+14 \gamma  \overline{r}^8-24 \beta  \overline{r}^6+48 \alpha  \overline{r}^4-128 \overline{r}^2+768\right) $  & NS & Fig.~\ref{K_Trirhm_xdot3_fig}(a) & S,~U,~S,~U,~S & rh\\
&&&  $(2.402 \times 10^{-5},~3.058 \times 10^{-8} )$ & 3.75166,~3.80059,~6.63736,~20.9299,~26.6688& yt\\

&&&&&\\
$- (1-\dot{x}^2+\alpha \dot{x}^4-\beta \dot{x}^6+\gamma \dot{x}^8-\delta \dot{x}^{10}) \dot{x}$ & $\overline{r} \left(-231 \delta  \overline{r}^{10}+252 \gamma  \overline{r}^8-280 \beta  \overline{r}^6+320 \alpha  \overline{r}^4-384 \overline{r}^2+512\right)  $  & U &  Fig.~\ref{R_Trirhm_xdot1_fig}(a) & S,~U,~S,~U,~S& hm\\
&& & $(2.544 \times 10^{-4},~6.62 \times 10^{-7} )$ & 1.77779,~1.82091,~2.86779,~12.5239,~15.7377& ic\\

&&&&&\\
$- (1-\dot{x}^2+\alpha \dot{x}^4-\beta \dot{x}^6+\gamma \dot{x}^8-\delta \dot{x}^{10}) \dot{x}^3$ & $\overline{r}^3 \left(-429 \delta  \overline{r}^{10}+462 \gamma  \overline{r}^8-504 \beta  \overline{r}^6+560 \alpha  \overline{r}^4-640 \overline{r}^2+768\right) $   & NS &  Fig.~\ref{R_Trirhm_xdot3_fig}(a)& S,~U,~S,~U,~S &\\
&&&  $(4.334 \times 10^{-4},~1.815 \times 10^{-6} )$ & 1.66034,~1.70743,~3.0214,~9.27171,~12.5056&\\
&&&&&\\
\hline
\end{tabular}}\\
\end{adjustbox}
\\ \\
S=Stable, U=Unstable, NS=Neutrally Stable, FP=Fixed Point

\noindent
{\bf Acknowledgement}

\noindent
{Sandip Saha acknowledges RGNF, UGC, India for the partial financial support. D S Ray is thankful to DST(SERB) for partial financial support under J C Bose National Fellowship. SS is thankful to Nikhil Pal (Visva-Bharati University) for useful discussions. We thank the anonymous reviewers for their careful reading of our manuscript and their many insightful comments and suggestions.} 

\bibliographystyle{unsrt}
\bibliography{References-Tri-rhythmicity_CNSNS-D-19-01667_Revised}{}

\end{document}